\documentclass {article}
\usepackage[dvips]{graphicx}
\bigskip \bigskip
\begin{document}

\bigskip \bigskip

\font\tenbi=cmmib10
\font\sevenbi=cmmib7
\font\fivebi=cmmib5

\textfont4=\tenbi
\scriptfont4=\sevenbi
\scriptscriptfont4=\fivebi
\def\bmit{\fam4}

\def\x{\times}
\def \op#1{\hbox {\boldmath $#1$}}

\font\hrm=cmr8
\baselineskip18pt
\parindent16pt
\parskip12pt

\centerline{\bf BOUNDARY CONDITIONS FOR A CAPILLARY FLUID}
\centerline{\bf IN CONTACT WITH A WALL}
\bigskip\centerline{Henri GOUIN \footnote{
 If you have read this paper and wish to be included in a mailing list that I maintain on the subject, then send e-mail to: \\ \centerline{henri.gouin@univ-cezanne.fr}
}(*)  \ and Witold KOSIN\'SKI (**)}
\bigskip
\noindent (*) Laboratoire de Mod\'elisation en M\'ecanique et Thermodynamique,\\ Facult\'e des Sciences,
Universit\'e d'Aix-Marseille, Case 322,\\
Avenue Escadrille Normandie-Niemen 13397 Marseille Cedex 20, FRANCE

\noindent (**) Institute of Fundamental Technological
Research,\\ Polish Academy of Sciences,
\'Swietokrzyska 21, 00-049 Warsaw, POLAND

\bigskip \bigskip
\centerline{\bf Abstract}\bigskip
Contact of a fluid with a solid or an elastic wall is investigated.
The wall exerts \emph{molecular forces}  on
the fluid which is locally
strongly nonhomogeneous. The problem is
approached
with a fluid energy of the second gradient form and a wall
surface energy depending on the value of the fluid density at the contact.
From the virtual work
principle are obtained limit conditions taking into account the fluid density,
its normal derivative to
the wall and the curvature of the surface.

\bigskip
{\bf 1. Introduction}\par The phenomenon of surface wetting is
 a subject of many experiments [1].
Such experiments have
been used to determine many important properties of the
wetting behavior for liquid on
 low energy surface [2].
In fact the wetting
transition of fluids in contact with solid surfaces is an important
field of research both for mechanics
and physical chemistry.
In the recent paper [3], the first author using  statistical methods proposed an explicit
 form for the energy of
interaction between solid surfaces and liquids. This energy yields
 a bridge connecting statistical mechanics and continuum mechanics.
 To obtain
boundary conditions between fluid and solid it is also necessary
 to know the behaviour of the fluid as well as the solid.

  We propose  a mechanical
model similar to that used in the mean-field theory of capillarity that leads to
the second gradient theory of continuous
media in fluid mechanics [4].
The theory is conceptually more straightforward than the Laplace
one to build a model of capillarity [5,6]. That theory takes
into account systems in which fluid interfaces are present [7].
The internal capillarity is one of the simplest cases since we are able to
calculate the superficial tension in the case of thin
interfaces as   well as in thick ones [8].
It is possible to obtain the
nucleation of drops and bubbles [9].

It seems that the approximation of the mean-field
theory is too simple to be quantitatively
accurate. However, it does provide a qualitative understanding. Moreover,
the point of  view, that the fluid in interfacial region may
be treated as a bulk phase with a local free energy density
and an additional
 contribution arising from the nonuniformity which may be
approximated by a gradient expansion truncated at the second order,
is most likely to be successful and perhaps even quantitatively
accurate near the critical point [10].

In this paper we connect both the interaction of a solid surface
 and a fluid phase by means of the
virtual work principle. The distribution of
fluid energy in the volume and the surface density energy on the solid
surface yield  boundary conditions. The conditions are different
from those obtained for a classical fluid within the theory of  gas dynamics.
We obtain an embedding effect for the
density of the fluid; moreover,  the conditions take  into account
 the curvature of the surface.
The result is extended to the case of an
elastic wall.

\noindent A discussion is obtained depending on the value of the density of
the fluid at the surface.

\noindent
Let us use asterisk "*" to denote  $conjugate$
(or  $transpose$) mappings
 or covectors (line vectors). For any vectors \ ${\bf
a}, {\bf b}$ \ we shall use the notation \ ${\bf a}^* {\bf b}$\  for their
 \ $ scalar \ product $
(the line vector is multiplied by the
  column vector) and \
\ ${\bf a} {\bf b}^*$\ or $\textbf{a}\otimes \textbf{b}$ for their
 \ $ tensor \ product $
 (the column vector is multiplied by the line vector).
The product of a mapping \ $A$ \ by a vector \ ${\bf a}$\ is denoted by \
$A\  {\bf a} $. Notation \ ${\bf b}^* \ A $ \ means
 covector
\ ${\bf c}^* $ \ defined by the rule \ ${\bf c}^* = ( A^*\  {\bf b} \ )^*$.
The {\rm div}ergence of a linear transformation \ $A$ \ is the covector $\ {\rm {\rm div}} A\ $ such that, for
any constant vector \ ${\bf a}, $
$${\rm {\rm div}} (A)\ {\bf a} \  = \ {\rm {\rm div} }\ (A\ {\bf a} \  ).$$
If $f(\bf x)$ is a scalar function of the vector $\bf x$ associated with the Euler variables in the
physical space, $ \displaystyle \partial f \over \displaystyle \partial {\bf x} $  is the linear form associated
with the gradient of $f$ and $\displaystyle  \partial f \over{\displaystyle  \partial  x^i}$ $=$  $\displaystyle ({\displaystyle\partial f
\over\displaystyle \partial {\bf x} })_i $. Consequently,   $ \displaystyle ({\partial f \over \partial {\bf
x}})^* $ $   = {\rm grad} \  f $

\bigskip
{\bf 2. Continuous mechanical model of capillary layers}

\bigskip
We consider a fluid in contact with a solid. The fluid occupies
the domain $\  D\ $ and its
boundary $\  \Sigma   \ $ which is common with the solid wall.
 Physical experiments prove that
the fluid is nonhomogeneous in the neighbourhood of $\  \Sigma
 $ [10]. It is  also possible
to consider
the fluid as a continuous medium by taking into account a
 \emph{capillary layer}  existing in
 the vicinity of
$\  \Sigma  $ and  a form of its stress tensor [11]. One way
 to present the behaviour of such a
fluid is to consider the specific internal energy
 $\ \varepsilon \ $ as a function of the
density $\ \rho\
$ as well as  $\displaystyle\ {{\rm grad}}\ \rho.$
Such an expression is known in continuum mechanics as {\it internal capillary energy}, see [4,5].
It is related to molecular models of strongly non homogeneous fluids in the frame of the mean field
theory and is equivalent to the van der Waals model of capillarity (see the review by Rowlinson and
Widom [10]). The energy $\ \varepsilon \ $ is also a function of the specific entropy. In the case of
isothermal media  at a given temperature, the specific internal energy is replaced by  the specific
free energy. In the mechanical case the entropy or the temperature are not concerned
by the virtual variations of the
medium. Consequently,  for an isotropic fluid, it is assumed that
$$\varepsilon\ =\ f(\rho , \beta ) $$
where  $\displaystyle\ \beta\ =\ (\overrightarrow{{\rm grad}} \ \rho)^2\ =
\ {\rm grad} \rho\ . \ {\rm grad} \rho\  $
(the dot . denotes the scalar product). The fluid is
submitted to external forces
represented by a  force potential
 $ \, \Omega\ $  per unit mass as a function of Euler 
variables $ {\bf x}$.

\noindent We denote by $\displaystyle\    {\bf x} \in \Sigma
\longrightarrow\ B({\bf x}) \in R $  the surface
density of energy of the solid wall. The total energy
$\ E\ $ of the fluid in $\ D\ $ and its
boundary $\ \Sigma \ $ is the sum of the three potential energies:
 $\displaystyle\ E\ =\
E_f+E_p+E_S\ $
with,
$$E_f\ =\ \int\int\int_D\rho\,\varepsilon (\rho , \beta)\ dv \ ,
 \  E_p\ =\
\int\int\int_D\rho\, \Omega({\bf x}) \ dv \  , \  E_S\ =
\ \int\int_\Sigma  B\ ds  $$
Let us denote by $\ \delta \ $ a variation of the position of the fluid as in [12].
The variation is
associated with the {\it virtual displacement}
  $$\displaystyle\  \textbf{x} \in  D\rightarrow \delta {\bf
x} = {{\mathbf \zeta}(x)} $$
We have the following results presented in Appendix,
$$\delta E_f=\int\int\int_D(-\, {\rm div} \  {\sigma})\, . \ {\bf \zeta}\ dv\ + $$ $$\int\int_\Sigma
 \left\{\matrix{\displaystyle -A {\
{d{\bf \zeta}_n}\over dn}}\ +\pmatrix{\displaystyle {2A\over R_m}\
 {\bf n} + {\rm grad} _{tg}A+ \sigma {\bf n}} .\ {\mathbf \zeta}\right\}
ds \eqno{(1)}$$
with
 $$\displaystyle\ \sigma=  - P {\bf I} - C \ {\rm grad} \rho \otimes
{\rm grad} \rho = - P {\bf I} - C ({\partial  \rho\over \partial {\bf x}} )^*\
{\partial \rho\over \partial {\bf x}} $$
where $ C  = 2\rho\varepsilon'_{\beta}$ and
 $P = \rho^2 \varepsilon'_\rho- \rho \   {\rm div} (C\
{\rm grad}\
\rho) $,
\\
\noindent    $\displaystyle \varepsilon'_\rho $ denotes the partial derivative of
$\varepsilon$ with respect to $ \rho $,
 ${{\zeta}}_n={\bf n}^*{{\mathbf \zeta}}\ $
where $\ {\bf n}\ $ is the external unit normal to $ \Sigma $ and
 $A=C\rho\ \displaystyle{d\rho\over dn}\ $  where  $\displaystyle\
{d\rho\over dn}=\ {\partial \rho\over \partial {\bf x}}\ {\bf n}.\ $

\noindent The scalar $
R_m\ $ is the mean curvature of $\Sigma $ and $\ {\rm grad}_{tg}\
$ is the tangential part of $\ {\rm grad}\ $ relatively
to
$\Sigma.$
\\
\noindent Moreover,
$$\delta E_p=\int\int\int_D\ \rho\ {\partial \Omega\over \partial
 {\bf x}} \, {{\mathbf \zeta}}\ dv\ = \int\int\int_D\ \rho\
{\rm grad}\,  \Omega\, .\,{{\mathbf \zeta}}\ dv \eqno{(2)}$$
and using the results presented in Appendix,
$$\delta E_S=\int\int_\Sigma\delta B-\pmatrix{\displaystyle\
{2B\over R_m}\,{\bf n} + {\rm grad}_{tg}B }.\,{{\mathbf \zeta}}\ ds \eqno{(3)} $$
One assumes that the volume mass in the fluid has a limit value $\ \rho_s $ at the wall $\Sigma$.
One assumes also that $\ B\ $ is a function of $\ \rho_s $ only.  These hypotheses are confirmed by
results presented in [3]. Then  $$ \matrix{\delta B = B'(\rho_s)\delta\rho_s  =  -\rho_s
B'({\rho_s})\ {\rm div} \, {\mathbf \zeta}} $$ Let us denote $\displaystyle\ G =-\rho_s  B'_{\rho_s}.\ $
Consequently, $$ \int\int_\Sigma \delta B\ ds =\int\int_\Sigma G\ {\rm div}\, {{\mathbf \zeta}}\ ds
=\int\int_\Sigma\ (\matrix{G\displaystyle\ {d{{ \zeta}_n}\over dn} - \ {2G  \over R_m}{\bf
n}\, .\ {{\mathbf \zeta}}-{\rm grad}_{tg} G \, .\ {{\mathbf \zeta}} })\ ds
$$
(see Appendix).

\noindent Now, $\displaystyle\ H = B(\rho_s)-\rho_s B'_{\rho_{s}} (\rho_s)\ $ is the Legendre
transformation of $\ B\ $ with respect to
$\
\rho_s$. Then,
$$\delta E_S=\int\int_\Sigma\ G\ {d{{\mathbf \zeta}}_n \over dn} - (\matrix{2H\displaystyle\ {{\bf
n} \over R_m}+{\rm grad}_{tg}H}) \, .\, {{\mathbf \zeta}}\ ds\eqno{(4)}$$

{\it The d'Alembert-Lagrange principle of virtuals works is expressed in the form} [12] :
$$\forall \ {\bf x}\in D  \rightarrow {{\mathbf \zeta} } ({\bf x})  ,\ \ \ \delta
E=0\eqno{(5)} $$
 Consequently, from the fundamental lemma of variation calculus, we obtain the balance
equation in the fluid $\ D\ $ and the boundary conditions on the solid wall $ \Sigma  $.

\noindent {\bf Equilibrium equations :}

\noindent From any arbitrary variation   $\displaystyle\    {\bf x} \in  D\rightarrow {\bf  \zeta}
({\bf x})\ $ such that  $\displaystyle\ {{\mathbf \zeta}} ={\bf 0}\ $ on  $\displaystyle\
 \Sigma  ,\ $ we take first
$$\int \int \int _D\pmatrix{\rho\displaystyle\
{\partial \Omega\over \partial {\bf x}}- {\rm div}\, \sigma }{{\mathbf \zeta}} \ dv=0$$
Consequently,
$$-\, {\rm div}\, \sigma +\rho \, {\partial \Omega\over \partial {\bf x}}=0 \eqno{(6)}$$
This equation is the well known equilibrium equation  [5,7,9]

\noindent {\bf Boundary conditions :

a) Case of a rigid (undeformed) wall.}

\noindent We consider a rigid wall. Consequently, the virtual
 displacements satisfy on $\Sigma\ $ the
condition  $\displaystyle\ {\bf n}^*  \ {{\mathbf \zeta}} =0$\,.
 Then, at the rigid wall
$$ \int\int_\Sigma (G-A)\displaystyle\ {d{{\zeta}}_n\over dn}+\left\{
\matrix{\displaystyle\ {2(A-H)\over R_m}{\bf n} +
{\rm grad}_{tg}(A-H) + \sigma  {\bf n}}\right\}  .\   {{\mathbf \zeta}}\ d\sigma=0 $$
\noindent Hence, we deduce the boundary conditions at the rigid wall
$$ \hbox{ For} \displaystyle\ {\bf x}\in  \Sigma ,\
 G-A=0\eqno{(7)}$$
and moreover, there  exists a Lagrange multiplier  $\displaystyle\  {\bf x} \in \Sigma  \rightarrow
\lambda({\bf x}) \in R\ $ such that 
$$ \ {2(A-H)\over R_m}\, {\bf n}  + {\rm grad}_{tg}(A-H)+ \sigma \,{\bf n} =\lambda \ {\bf
n} \eqno{(8)} $$

{\bf b) Case of a elastic (non-rigid) solid wall.}

\noindent In such a case the equilibrium equation(6) is unchanged.
On $\Sigma\ $ the condition (7) is also unchanged.
The only different condition comes from the fact that we do not have anymore the slipping
condition for the virtual displacement
  $(\displaystyle  {\bf
n}^*{{\mathbf \zeta}}=0).\ $

\noindent Due to the possible deformation of the wall, the virtual  work of mechanical
 stresses on $\
\Sigma \ $ is
$$\delta E_e=\int\int_\Sigma{\bf T}^*{{\mathbf \zeta}}\ ds $$
with  $\displaystyle\ {\bf T} \ =\   Q\ {\bf n} $ representing the stress (loading) vector,
 where $\
Q\ $ is the value of the Cauchy stress tensor of the wall on the boundary $\, \Sigma $.

\noindent Relation (8) is remplaced by :
$$2\ {(A-H)\over R_m}\,{\bf n} + {\rm grad}_{tg}(A-H)+ \sigma \,{\bf n} = -{\bf T} \eqno{(9)}  $$

\bigskip
\bigskip
{\bf 3. Analysis of the boundary conditions}

\noindent Relation (7) yields :
$$ C\ {d\rho\over dn}+B'_\rho=0\eqno{(10)}$$
and we obtain,
  $$\displaystyle\ H - A = B.\ $$
Consequently, from the definition of $\sigma $,
 $$\displaystyle\  \sigma \ {\bf n} = P\,{\bf n} - C {d\rho\over dn}\ {\rm grad} \rho.\ $$
The tangential part of equation (8) is always verified and equation (8) yields the value
of the Lagrange multiplier $\ \lambda$.

 \noindent For an elastic (non-rigid) solid wall we obtain
$$ T^*_{tg}=0\hskip 1 cm \hbox{ and }\hskip 1 cm T_n\ =\ {2B\over R_m}+P-B'_\rho\,
{d\rho\over dn}\eqno{(11)}  $$
where $\ T^*_{tg}\ $  and $\ T_n\ $ are respectively the tangential and the normal components of $\, {\bf T}$. Taking into account equation (10),  $\displaystyle\ T_n=P+\ {2B\over R_m}+\ {1\over
C}(B'_\rho)^2  $  and equations (11) yield the value of the stresses in the elastic (non rigid)
medium. The only new condition comes from equation (10).

We have the consequences:

\noindent In [3]  we propose the surface energy in the form  $\displaystyle\ B(\rho)=-\gamma _1
\rho  +\ {\gamma _2\over 2}\rho^2\  $  with $\ \gamma_1 ,\ $ and  $\
\gamma_2\
$ as two positive constants. We obtain the condition for the fluid density on the wall
$$C\ {d\rho\over dn}=\gamma _1-\gamma _2\rho\eqno{(12)}  $$
Denoting again $\, \rho_s\, $ the value of $\ \rho \ $ on the wall $\, \Sigma \, $ (the limit value of
the volumic mass of the fluid on the wall $\, \Sigma\ $),  we obtain that
  $\displaystyle  \, {d\rho \over dn}\, $ is
positive (or negative) in the vicinity of the  wall if
$\displaystyle\ \rho_s<\rho_i\ $ (or  $\displaystyle\ \rho_s>\rho_i\ $ )
with $\displaystyle\ \rho_i=\ {\gamma _1\over \gamma _2}$ which
is the {\it bifurcation fluid density} at the wall.

\noindent If  $\displaystyle\ \rho_s<\rho_i\ $  we have a lack of fluid density at the wall.
If  $\displaystyle\ \rho_s>\rho_i\ $  we have a excess  of fluid density at the wall.

{\bf Conclusion}

For conservative  medium, the first {\rm grad}ient theory corresponds to the case of compressibility. To
take  into account superficial effects acting between solids and fluids,
 we propose to use the model of fluids endowed with
capillarity.  The theory interprets
the capillarity in a continuous way
and contains Laplace's theory. The model corresponds for solids to \emph{elastic materials with
couple stresses} indicated by Toupin in [13].
\\
We notice that the extension to the dynamic case is straightforward: by virtual work
principle, equation (6) takes the form:
$$\rho \gamma^* - {\rm div}\, \sigma+\rho\,  {\partial \Omega\over \partial {\bf x}}=0,
$$
where $\gamma$ denotes the acceleration of the fluid. Equations (10), (11), (12) and consequences in
paragraph 3  are unchanged.

{\bf Acknowledgment}

The authors are grateful to the Polonium Program of Co-operation $N^0 \  7075$ between Polish KBN and
French Foreign Office for the financial support of this research.

{\bf  References}

[1] A. E. Van Giessen, D. J. Bukman, B. Widom, Contact angles of liquid drops on low-energy solid surfaces, J. Colloid
Interface Sci., {\bf 192}, 257-265, 1997.

[2] J. W. Cahn, Critical point wetting, J. Chem. Phys., {\bf 66},  3667-3672, 1977.

[3]  H. Gouin, Energy of interaction between solid surfaces and liquids, J. Phys. Chem., {\bf 102},
1212-1218 , 1998.

[4] P. Germain, La m\'ethode des   puissances virtuelles
en m\'ecanique des milieux continus, J. de M\'ecanique, {\bf 12},
235-274, 1973.

[5] P. Casal, La th\'eorie du second {\rm grad}ient et la capillarit\'e,
Comptes Rendus Acad. Sc. Paris,
{\bf 274},  1571-1574, 1972.

[6] P. Casal, H. Gouin, Sur les interfaces liquide-vapeur non isothermes,
 J. de M\'ecanique Th\'eorique et Appliqu\'ees,
{\bf 7},  689-718, 1988.

[7] H. Gouin, Utilization of the second {\rm grad}ient theory in continuum mechanics
 to study the motion and thermodynamics of
liquid-vapor interfaces, Physicochemical Hydrodynamics -
Interfacial Phenomena, B, {\bf 174},  667-682, 1987.

[8] E. Dell'Isola, W. Kosi\'nski, Deduction of thermodynamic balance law for bidimensional
nonmaterial directed continua modelling
interface layers  , Arch. Mech. {\bf 45}, (3), 333-359, 1993.

[9] F. dell'Isola, H. Gouin, G. Rotoli, Nucleation of spherical shell-like
interfaces by second {\rm grad}ient theory:
numerical simulations, Eur. J. Mech. B / Fluids, {\bf 15},
 545-568, 1996.

[10] J. S. Rowlinson, B. Widom, Molecular theory of Capillarity,
Clarendon Press, Oxford, 1984.

[11] P. Seppecher, Equilibrium of a Cahn-Hilliard fluid on a wall:
influence of the wetting properties of the film upon
the stability of a thin film, Eur. J. Mech. B / Fluids, {\bf
12},  69-84, 1993.

[12] J. Serrin, Mathematical principles of classical fluid mechanics,
 Encyclopedia of Physics, Flugge S. Ed., vol. VIII/1,
Springer-Verlag, Berlin,  1959.

[13] R.A. Toupin, Elastic materials with couple stresses, Arch. Rat. Mech. Anal.,
{\bf 11}, 385-398, 1962.

\bigskip \bigskip
\centerline{{\bf Appendix}}
\bigskip
\noindent First of all  we recall the following fact issued from  differential geometry:\\
Let $\ \Sigma\ $ be a surface in the 3-dimensional space and $\ {\bf n}\ $ its external normal.
 For any vector field $\ {{\mathbf \zeta}},\ $
$$ {\bf n^*}  rot ({\bf n}\x {{\mathbf \zeta}}) = {\rm div} \ {{\mathbf \zeta}}+\
{2\over R_m}\   {\bf n^*}  {{\mathbf \zeta}} - {\bf n^*} \ {\partial
  {{\mathbf \zeta}}\over \partial {\bf x}}\
 {\bf n}.
$$
Then, for any scalar field $\ A,\ $ we obtain:
$$\matrix{A \ {\rm div}\ {{\mathbf \zeta}}=A\ \displaystyle{d{{ \zeta}}_n\over dn}-\
{2A\over R_m}\,{{ \zeta}}_n-({\rm grad}^*_{tg}A)   \, {{\mathbf \zeta}}+
{\bf n}^*\ rot\ (A{\bf n}\x
{{\mathbf \zeta}})\cr\cr
={\rm tr}\left[\matrix{ \pmatrix{\ {\displaystyle \partial A\over \displaystyle \partial
{\bf x}}({\bf n}{\bf n}^*-{\bf 1})
  -\displaystyle\ {2A\over R_m}\ {\bf n}^*}{{\mathbf \zeta}}  }\right ]+A\
\displaystyle{d{{\zeta}}_n\over dn}\ +{\bf n}^* rot (A{\bf n}\x {{\mathbf \zeta}})
}\eqno{(A.1)}$$

\noindent
Let us calculate  $\displaystyle\ \delta E_f\ $: since $D$ is a material volume,
$$E_f=\int\int\int_D\rho\  \varepsilon\ dv\ \ \ \ \ \Rightarrow \ \ \ \ \delta E_f=
\int\int\int_D\rho\ \delta\varepsilon\ dv$$
with  $\displaystyle\ \delta\varepsilon=\ {\partial \varepsilon \over
\partial \rho}\ \  \delta\rho+\
{\partial \varepsilon\over \partial \beta}\  \delta\beta  $.

\noindent
From  $\displaystyle\ \delta\ {\partial \rho\over \partial {\bf x}}=\
{\partial \delta\rho\over
\partial {\bf x}}-{\partial \rho\over \partial {\bf x}}\ {\partial
{{\mathbf \zeta}}\over \partial {\bf
x}},\  \ $  we deduce
$$\rho\,\varepsilon'_\beta\, \delta\beta=2\rho\, \varepsilon'_\beta\ \delta (
{\partial \rho\over \partial
{\bf x}})\ {\partial \rho\over \partial {\bf x}} ^*
 = C \pmatrix{\displaystyle  {\partial
\delta\rho\over
\partial {\bf x}}-\ {\partial
\rho\over \partial {\bf x}}\ {\partial {{\mathbf \zeta}}\over \partial {\bf x}}  }
\displaystyle{\partial \rho\over \partial {\bf x}}^*
 $$
with  $\displaystyle\ 2\rho \,\varepsilon'_\beta=C.\ $

\noindent In the mean-field molecular theory, the quantity $\, C\, $ is assumed constant [10],
 but it is not necessary for our calculations. One can
suppose the scalar $\, C\, $ is a general
 function
of $\
\rho\ $ and even $\ \beta$.
Then
$$\rho\,\varepsilon'_\beta\ \delta\beta = {\rm div} (C\ {\rm grad}\ \rho\ \delta\rho) -
{\rm div}(C\ {\rm grad}\ \rho)\,
\delta\rho-tr\pmatrix{C\ {\rm grad} \rho \ \
{\rm grad}^*\rho\displaystyle\ {\partial {{\mathbf \zeta}}\over \partial {\bf x}}\ }
$$
Due to the fact that   $\displaystyle\ \delta\rho=- \rho\, {\rm div}\ {{\mathbf \zeta}}\
 $ (see [12]),
$$
\matrix{\rho\, \delta\varepsilon={\rm div} (C\ {\rm grad}\ \rho\ \delta\rho)-
\Big(\rho^2\, \varepsilon'_\rho - \rho \ {\rm div}(C\
 {\rm grad}\ \rho\ )\Big){\rm div} \ {{\mathbf \zeta}}\cr\cr
-\,{\rm div} (C\ {\rm grad}\,  \rho\ \, {\rm grad}^*\rho\hskip 0.2cm{{\mathbf \zeta}})+{\rm div}(C \ {\rm grad} \, \rho\ \,
{\rm grad}^*\rho) \, {{\mathbf \zeta}}}
$$
$$
\matrix{\rho\,\delta\varepsilon={\rm div}  \Big ( C\ {\rm grad}\, \rho \ \delta\rho -
(C\ {\rm grad}\ \rho\ {\rm grad}^*\rho){{\mathbf \zeta}}-P{{\mathbf \zeta}}\Big )}
$$
$$
+\displaystyle\ {\partial P\over \partial {\bf x}}\,{{\mathbf \zeta}}+{\rm div} (C\ {\rm grad} \
 \rho\ {\rm grad}^*\ \rho)\, {{\mathbf \zeta}}
$$
Then
$$\displaystyle\delta
E_f=\int\int\int_D\left(\matrix{\displaystyle\
 {\partial P\over \partial {\bf
x}}+{\rm div}(C\ {\rm grad} \, \rho\ {\rm grad}^* \rho)}\right){{\mathbf \zeta}}\,  dv$$
$$\displaystyle\ -\ \int\int\int_D {\rm div} \Big( C\rho\ {\rm grad}\ \rho\ {\rm div} \ {{\mathbf \zeta}}+ C\ {\rm grad} \
\rho\ {\rm grad}^*\rho \ \,{{\mathbf \zeta}}+ P{{\mathbf \zeta}}\Big)
 \ dv$$
 $$\displaystyle\ =\int\int\int_D -({\rm div}\,\sigma)\,{{\mathbf \zeta}}\
dv+\int\int_\Sigma(-A\ {\rm div}\, {{\mathbf \zeta}}+{\bf n}^*\sigma \ {{\mathbf \zeta}})\, ds $$
Taking into account (A.1), we deduce immediately
$$\displaystyle \delta E_f= \int\int\int_D-({\rm div}\, \sigma)\,{{\mathbf \zeta}}\ dv$$
$$\displaystyle+\int\int_\Sigma \Big(-A\displaystyle\ {d{{ \zeta}}_n\over
dn}+ ({\displaystyle {2A\over R_m}\,{\bf n}^*+{\rm grad}^*_{tg}A+{\bf n}^*\sigma })\,
{{\mathbf \zeta}}  \Big )ds
 \displaystyle +\int\int_\Sigma{\bf n}^* {\rm rot}(A\ {\bf n}\x{{\mathbf \zeta}})
 ds  $$
But   $\displaystyle\ \int\int_\Sigma{\bf n}^*{\rm rot}(A\ {\bf
n}\x{{\mathbf \zeta}})\,ds =
\int_\Gamma A{\bf t} \, .\, ({\bf n}\x {{\mathbf \zeta}}) \ d\ell  $
$
 = \int_\Gamma  A ({\bf t, n,}\,{{\mathbf \zeta}}) \ d\ell $

\noindent where $\ \Gamma\ $ is the line boundary of $\, \Sigma $ and ${\bf t}$ its tangent unit
vector.
 If $\ {\bf n}'={\bf t}\x{\bf n}\ $ we obtain the relation
$$ \displaystyle\delta E_f=\int\int\int_D(- {\rm div}\, \sigma)\,{{\mathbf \zeta}}\ dv\ +
$$
$$
\int\int_\Sigma \Big(-A\displaystyle\ {d{{ \zeta}}_n\over
dn}+ ({\displaystyle  {2A\over R_m}\,{\bf n}^*+{\rm grad}^*_{tg}A+{\bf n}^*\sigma })\,
{{\mathbf \zeta}}  \Big )ds +\int_\Gamma A\,{\bf n'}^*{{\mathbf \zeta}}\ d\ell
\eqno{(A.2)}$$
In the following  we assume that $\, \Sigma\,  $
has no boundary and consequently, the term associated with $\ \Gamma\ $ vanishes.
\\
Let us calculate $ \ \delta E_S$
$$E_S=\int\int_\Sigma\ B\ ds $$
Then
$$\displaystyle\ \delta E_S=\int\int_\Sigma\ \delta B\ -\Big ( {\bf
n}^*\displaystyle\, {2B\over R_m }
+{\rm grad}^*B\,({\bf 1}-{\bf nn}^*) \Big)\,{{\mathbf \zeta}} \ ds
+\int_\Gamma A\,{\bf n'}^*{{\mathbf \zeta}} \ {d\ell}   \eqno (A.3)$$
We notice that $\ {\rm grad}^*B({\bf 1}-{\bf nn}^*)\ $ belong to the
 tangent plane to $\ \Sigma$.
\\
Let us proof equation (A.3):  If we write  $\displaystyle\ E_S=\int\int_\Sigma B\ \det\
 ({\bf n},d_1{\bf x},d_2{\bf x})\, $, where  $\displaystyle\ d_1{\bf x} \ $ and
$\displaystyle\ d_2{\bf x}\
$ are the coordinate lines of $\ \Sigma,\ $ we may write,
$$ E_S=\int\int_{\Sigma_0}B\ \det\ F\ \ \hbox{det} \ (F^{-1}{\bf n},d_1{\bf
X},d_2{\bf X})$$
where $\ \Sigma_0\ $ is the image of $\, \Sigma \,$ in a reference space in Lagrangian
coordinates  $\displaystyle\ {\bf X}\ $  and $\ F\ $ the deformation gradient tensor
$\displaystyle  \, {\partial {\bf x}\over \partial {\bf X}}.$

\noindent Then,  $$\displaystyle\ \delta E_S=\int\int_{\Sigma_0}\delta B\ \det\ F\
\hbox{det}\ (F^{-1}{\bf n},d_1{\bf X},d_2{\bf X})\ +$$
$$\displaystyle\ \int\int_{\Sigma_0}B \, \delta   \Big (\det\ F\ \hbox{det}\
(F^{-1}{\bf n},d_1{\bf
X},d_2{\bf X})\Big )  $$
Moreover,
$$\displaystyle\ \int\int_{\Sigma_0}B\,  \delta   \Big ({\rm  det}\, F\ \hbox{det}\
(F^{-1}{\bf n},d_1{\bf
X},d_2{\bf X})\Big ) = $$
$$\int\int_\Sigma B\ {\rm div}\, {{\mathbf \zeta}} \   \det\,
 ({\bf n},d_1{\bf x},d_2,{\bf x})\ +  B\,  \det\, \big (\displaystyle
 {\partial {\bf n}  \over \partial {\bf x}}{{\mathbf \zeta}},d_1{\bf x},d_2{\bf x}\big )
-  B\ \det\, \big (\displaystyle\
 {\partial{{\mathbf \zeta}}  \over \partial {\bf x}}{\bf n},d_1{\bf x},d_2{\bf x} \big )$$
$$\displaystyle = \int\int_\Sigma\pmatrix{{\rm div} (B\,{{\mathbf \zeta}}  )-{\rm grad}^*
B \ {{\mathbf \zeta}}  -B\, {\bf n}^*\displaystyle\ {\partial {{\mathbf \zeta}}  \over
\partial {\bf x}} \ {\bf n} }ds$$
From (A.1) we obtain,
$$
{\rm div} (B\,{{\mathbf \zeta}}) - B \   ({\rm div}\,{\bf n})\ {\bf n}^*
{{\mathbf \zeta}} - {\bf n}^* {{\partial { B {\mathbf \zeta}}}\over
\partial {\bf x}}  \ {\bf n}
= {\bf n}^*\ {\rm rot}\ (B\,{\bf n} \x {{\mathbf \zeta}})
$$
Then,
$$\displaystyle\ \int\int_{\Sigma_0}B\,\delta   \big (\det\ F\ \hbox{det}\
(F^{-1}{\bf n},d_1{\bf
X},d_2{\bf X})\big ) = $$
$$\displaystyle\ \int\int_{\Sigma_0} \Big ( B \
 ({\rm div}\,{\bf n} ) \ {\bf n}^*+{\rm grad}^*B\,({\bf nn^*-1}) \Big ) {{\mathbf \zeta}}
\ ds+\int\int_\Sigma{\bf n}^*\ {\rm rot}\, (B{\bf n}\x{{\mathbf \zeta}}  )\,ds $$
and we obtain equation (A.3) with  $\displaystyle\ {\rm div} \ {\bf n}=-\ {2\over R_m}. $

\noindent We assume that $\, \Sigma\, $ has no boundary and consequently, the term
associated with $\ \Gamma\ $ is null.

\end{document}